\documentclass[10pt,conference]{IEEEtran}
\usepackage[T1]{fontenc}
\usepackage{lmodern,microtype,booktabs,tabularx,array,enumitem,xcolor,listings,amsmath,amssymb,url,needspace}
\usepackage[hidelinks]{hyperref}
\hypersetup{
  pdfauthor={Rishabh Mehan},
  pdftitle={Can MCP Clients Decide What to Do After Failure? A Result-Only Actionability Audit}
}
\definecolor{codebg}{HTML}{F4F6F8}
\definecolor{codeblue}{HTML}{155D8B}
\newcommand{\ControlDetect}{18}
\newcommand{\ControlPolicy}{8}
\newcommand{\ControlCause}{0}
\newcommand{\ControlTarget}{0}
\newcommand{\ControlExecutable}{0}
\newcommand{\ControlReplay}{0}
\newcommand{\ContentDetect}{21}
\newcommand{\ContentPolicy}{19}
\newcommand{\ContentCause}{19}
\newcommand{\ContentTarget}{18}
\newcommand{\ContentExecutable}{0}
\newcommand{\ContentReplay}{0}
\newcommand{\ControlJointDP}{8}
\newcommand{\ControlJointD}{10}
\newcommand{\ControlJointNone}{3}
\newcommand{\ContentJointDPCT}{18}
\newcommand{\ContentJointDPC}{1}
\newcommand{\ContentJointD}{2}
\newcommand{\ControlIncomparablePairs}{0}
\newcommand{\ContentIncomparablePairs}{0}

\newcommand{\LlamaCodeAnyFalseSuccessCount}{0}
\newcommand{\LlamaCodeStrictCount}{39}
\newcommand{\LlamaCodeCoverageCount}{105}
\newcommand{\LlamaCodeRetryCount}{1}
\newcommand{\LlamaCodeInvalidCount}{0}
\newcommand{\LlamaCodeAdditionalCalls}{1}

\newcommand{\LlamaNormalizedAnyFalseSuccessCount}{0}
\newcommand{\LlamaNormalizedStrictCount}{42}
\newcommand{\LlamaNormalizedCoverageCount}{105}
\newcommand{\LlamaNormalizedRetryCount}{1}
\newcommand{\LlamaNormalizedInvalidCount}{0}
\newcommand{\LlamaNormalizedAdditionalCalls}{1}

\newcommand{\LlamaProseAnyFalseSuccessCount}{0}
\newcommand{\LlamaProseStrictCount}{105}
\newcommand{\LlamaProseCoverageCount}{105}
\newcommand{\LlamaProseRetryCount}{0}
\newcommand{\LlamaProseInvalidCount}{0}
\newcommand{\LlamaProseAdditionalCalls}{0}

\newcommand{\LlamaStructuredAnyFalseSuccessCount}{0}
\newcommand{\LlamaStructuredStrictCount}{95}
\newcommand{\LlamaStructuredCoverageCount}{105}
\newcommand{\LlamaStructuredRetryCount}{0}
\newcommand{\LlamaStructuredInvalidCount}{0}
\newcommand{\LlamaStructuredAdditionalCalls}{0}

\newcommand{\QwenCodeAnyFalseSuccessCount}{2}
\newcommand{\QwenCodeStrictCount}{5}
\newcommand{\QwenCodeCoverageCount}{76}
\newcommand{\QwenCodeRetryCount}{24}
\newcommand{\QwenCodeInvalidCount}{6}
\newcommand{\QwenCodeAdditionalCalls}{48}

\newcommand{\QwenNormalizedAnyFalseSuccessCount}{5}
\newcommand{\QwenNormalizedStrictCount}{6}
\newcommand{\QwenNormalizedCoverageCount}{73}
\newcommand{\QwenNormalizedRetryCount}{24}
\newcommand{\QwenNormalizedInvalidCount}{5}
\newcommand{\QwenNormalizedAdditionalCalls}{48}

\newcommand{\QwenProseAnyFalseSuccessCount}{2}
\newcommand{\QwenProseStrictCount}{30}
\newcommand{\QwenProseCoverageCount}{77}
\newcommand{\QwenProseRetryCount}{23}
\newcommand{\QwenProseInvalidCount}{11}
\newcommand{\QwenProseAdditionalCalls}{42}

\newcommand{\QwenStructuredAnyFalseSuccessCount}{2}
\newcommand{\QwenStructuredStrictCount}{39}
\newcommand{\QwenStructuredCoverageCount}{84}
\newcommand{\QwenStructuredRetryCount}{26}
\newcommand{\QwenStructuredInvalidCount}{8}
\newcommand{\QwenStructuredAdditionalCalls}{55}

\newcommand{\TrajectoryCount}{840}

\title{Can MCP Clients Decide What to Do After Failure?\\
A Result-Only Actionability Audit}
\author{\IEEEauthorblockN{Rishabh Mehan}
\IEEEauthorblockA{Independent\\
\href{mailto:rishabhmehan@gmail.com}{rishabhmehan@gmail.com}}}

\begin{document}
\maketitle
\begin{abstract}
A client that receives \texttt{isError:true} knows that something went wrong. It may still have no machine-readable basis for deciding whether to fix an argument, authenticate, wait, choose another tool, or stop. This paper studies what deterministic software can learn from a \emph{completed MCP failure result alone}; request arguments, schemas, discovery history, authentication state, transport metadata, host policy, and application state are outside that boundary. We introduce a six-part actionability profile and apply it with record-level evidence. In a small illustrative study of 21 safely induced failures from ten reachable sampled servers, typed fields expose failure in \ControlDetect{} cases and a broad policy in \ControlPolicy{}, yet expose no specific cause, target, executable repair, or replay constraint. Prose often carries more cause and target information, at the price of making semantic interpretation part of the recovery path. A lexical source audit finds the same text-centered pattern. Finally, a fail-closed prototype demonstrates how a separate experimental control plane could support deterministic branching. The result is deliberately narrower than an ecosystem survey or an agent benchmark: completed MCP results often make failure observable, sometimes make a broad response possible, and rarely make concrete recovery or safe replay self-contained in this sample.
\end{abstract}

\section{Introduction}
A retry is not a neutral response to failure. Repeating a malformed request wastes a call; repeating an unauthenticated request does not restore a credential; repeating a partially committed operation may do real damage. Other failures call for a corrected argument, a delay, a different tool, or human intervention. A Boolean failure marker cannot distinguish among those choices.

MCP gives hosts, clients, and tool servers a common language for discovery and invocation \cite{mcp_spec_2026}. For an ordinary tool-execution failure, the JSON-RPC exchange can still succeed while the result says \texttt{resultType: complete}, sets \texttt{isError: true}, and supplies explanatory content. Protocol failures take a different route through typed JSON-RPC errors \cite{mcp_tools_2025,jsonrpc20}. The protocol also has a specialized \texttt{InputRequiredResult} and a multi-round-trip workflow for asking for missing input \cite{mcp_mrtr_2026}. Those mechanisms matter, but general completed failures still lack standard fields for cause, repair target, policy, timing, or replay safety \cite{mcp_schema_2026}. In other words, MCP specifies how to observe many failures more fully than it specifies how to recover from them.

Our question is therefore deliberately bounded: \emph{given only a completed MCP failure result, how far can deterministic software get toward a safe next action before it has to interpret server-specific prose or look elsewhere?} A production client can also consult the original request, the tool schema, discovery history, authentication state, transport metadata, host policy, and application state. We set those sources aside so that the result itself can be measured. The paper contributes:
\begin{enumerate}[leftmargin=*,itemsep=1pt]
\item a separately coded, non-cumulative \emph{error-actionability profile} and componentwise partial order;
\item an auditable, record-level method applied to a version-scoped specification, high-precision lexical corpus audit, and illustrative live case study; and
\item an executable fail-closed conformance refinement that separates machine controls from explanations.
\end{enumerate}
We also retain a small, exploratory policy-selection test with two local models. It is a supporting check rather than a central result. The 21 failures are enough to exercise the construct and demonstrate the boundary, but not to estimate registry-wide prevalence, raw-wire behavior, deployed-agent performance, or end-to-end recovery.

\section{Failure Contracts and Related Work}
\subsection{MCP failure surfaces}
MCP failures reach clients through several routes. A \emph{protocol error} is a JSON-RPC error with an integer code, message, and optional data. A \emph{tool-execution error} arrives inside an otherwise successful JSON-RPC response and marks its result with \texttt{isError: true}. HTTP and transport failures sit below both. We also observed a fourth, awkward case: a \emph{success-shaped error}, where the text says the operation failed but the typed result does not.

The SDK can alter which of these surfaces an application sees. Official TypeScript SDK guidance describes tool errors as \texttt{isError:true} results read by the model, tells server authors to put a recovery hint in text, and skips output-schema validation for those results \cite{mcp_ts_errors}. The Python SDK separates ordinary tool failures from protocol and low-level failures \cite{mcp_py_errors,mcp_py_lowlevel}. Accordingly, our dataset records the Python client's normalized view; we do not pretend that it preserves the wire representation. An open MCP issue now proposes a schema-governed structured error object for this path \cite{mcp_structured_error_issue}. We treat that issue as evidence that the design question is active, not as part of the standard.

\subsection{Machine-facing precedents}
Other protocols make this separation explicit. HTTP Problem Details provides a machine discriminator alongside human-facing detail and warns clients not to automate by parsing that detail \cite{rfc9457}; HTTP also defines \texttt{Retry-After} where delay matters \cite{http_semantics}. gRPC pairs a constrained status vocabulary with optional prose, then places retryable codes, attempt limits, backoff, and jitter in a separate retry configuration \cite{grpc_status,grpc_retry}. None of these mechanisms solves application-specific recovery. They do show that controls and explanations can be designed for different consumers.

That distinction is familiar in operations. Backoff, jitter, retry limits, and circuit breakers exist because an automatic retry can deepen an outage \cite{aws_backoff,circuit_breaker,google_cascading}. Retry decisions are themselves a recurring source of software defects \cite{retry_bugs}. This literature motivates host-enforced policy, while also setting a clear boundary around our own replay loop: returning the same stored error is not an outage experiment.

Structured recovery for MCP is not a blank slate. SERF proposes an \texttt{isError} envelope with categories, retryability, timing, ordered actions, and a deterministic recovery procedure; its recovery-rate claims remain hypotheses for a proposed synthetic evaluation \cite{serf_mcp}. AdCP 3.1.2 carries a namespaced \texttt{adcp\_error} through MCP \texttt{structuredContent}, JSON-RPC \texttt{error.data}, or guarded text fallback, and supplies recovery classes, retry bounds, extraction order, validation, and test vectors \cite{adcp_transport_errors}. We therefore do not claim to have discovered the gap or invented the control/content split. Our contribution is the narrower one: an auditable way to measure what a completed result makes actionable, together with a fail-closed conformance refinement exercised against record-level evidence.

\subsection{Agent failure benchmarks and policy systems}
Recent empirical work maps runtime and software faults in MCP repositories and practitioner reports \cite{mcp_runtime_fault_taxonomy,mcp_real_faults}; Winston and Just provide a broader taxonomy for tool-augmented LLM systems \cite{tallm_failure_taxonomy}. Those studies ask what fails and where. We ask a different question about the failure record: what can ordinary software safely infer from it?

Tool-use benchmarks usually begin later in the loop, with tool choice, arguments, or full trajectories \cite{apibank,toolllm,gorilla,toolsandbox}. \emph{Tools Fail} examines silent faulty outputs, and FAIL-TaLMs studies underspecified queries and unavailable tools \cite{tools_fail_silent,fail_talms}. ToolFailBench focuses on behaviors such as skipping tools, ignoring results, and fabrication \cite{toolfailbench}. AgentCheck is especially close: it injects 12 MCP fault types and evaluates five agents in a replay/intervention workbench with deterministic and human-validated scoring \cite{agentcheck}. Several newer benchmarks study dynamic recovery under injected failures, sometimes across multiple model families \cite{when_tools_fail,retry_switch_abstain,failing_tools,toolmisusebench,planbenchxl,toolbenchx}. Other work protects non-atomic calls with postconditions and idempotency, or moves recovery policy outside the model altogether \cite{verified_tool_calls,guardrails_infrastructure}.

Our unit of analysis sits upstream of those agent evaluations. Before asking whether an agent recovers, we ask what the completed MCP result gives a deterministic client to work with. The same profile can later be applied to benchmark failures as well as live observations.

\section{A Non-Cumulative Actionability Profile}
For an observation $o$, define
\[
\mathbf{a}(o)=(D,P,C,T,E,R)\in\{0,1\}^{6},
\]
where each coordinate is coded separately and no positive assignment is inherited:
\begin{description}[leftmargin=0pt,style=nextline,itemsep=2pt]
\item[$D$ --- Failure detection.] Software distinguishes failure from success.
\item[$P$ --- Coarse policy.] It can select among request repair, authentication, delayed retry, alternate selection, or stop/escalation.
\item[$C$ --- Cause/subtype.] It distinguishes causes such as unknown tool, invalid type/value, authentication, rate limit, or upstream failure.
\item[$T$ --- Repair target.] It identifies an offending tool, parameter, credential, or dependency.
\item[$E$ --- Executable repair.] It can construct or execute the recovery without guessing a missing value or performing another discovery step.
\item[$R$ --- Replay constraints.] Timing, unchanged-request safety, and relevant side-effect state are explicit.
\end{description}

The six coordinates should not be read as rungs on a ladder. A rate-limit response can provide timing ($R$) without a corrected request ($E$), and a target pointer ($T$) need not identify a precise cause ($C$). We compare observations by componentwise dominance: $o_1\succeq o_2$ iff every capability evidenced by $o_2$ is also evidenced by $o_1$. This permits incomparable profiles, although none occur in the present sample. We also set a deliberately high bar for $E$: knowing a repair procedure, possessing the replacement value, and having authority to act are collapsed into one end-to-end threshold. Because no record reaches $E$, the present data cannot tease those sub-capabilities apart.

We read each record twice. In the \emph{control} view, every string---even a JSON object serialized as text---is opaque. In the \emph{content-assisted} view, model-visible text may be interpreted. Thus \texttt{isError} earns $D$ in the control view but not $P$. JSON-RPC \texttt{-32602} earns only broad request repair: in this sample it covers both unknown tools and bad arguments, so it cannot identify the cause. Text that names a field may earn $T$ without earning $E$ if it supplies no valid replacement. The release gives an evidence pointer and rationale for every bit. The distance between the two views is practical: whenever prose supplies the missing information, recovery inherits the cost and fallibility of a semantic interpreter. That problem is separate from information the result omits but a client might obtain elsewhere.

More than one response can be safe. For each controlled scenario, we publish a set containing the cause-specific repair and stop/escalate. Since stopping is safe in all 21 cases, membership in that set says little about useful progress. The model stress test therefore reports preferred-policy matches alongside valid non-stop coverage; selective accuracy is left to the artifact as a sensitivity calculation.

\section{Illustrative Case Study}
\subsection{Specification, sample, and static candidates}
We froze the specification and schema available on August 19, 2026, using protocol revision 2026-07-28 as the reference point. The accompanying registry snapshot held 23,569 latest records, from which we derived 16,699 unique GitHub repositories \cite{mcp_registry_api,mcp_registry_about}. Seed 20260819 drew 54 repositories: 12 each from npm, PyPI, and remote-only listings, and six each from MCPB, OCI, and other listings. When five initial draws were inaccessible, the next seeded repository in the same stratum replaced each one. The public release preserves the derived frame, draw order, exclusions, replacements, final sample, and hashes. It can replay the sampling step from that frame, though it cannot reconstruct the omitted raw registry response.

We then used a deliberately narrow lexical scanner to find likely error construction. It excludes dependencies and builds, while retaining and labeling generated, test, and fixture paths for audit. All 16 \texttt{isError} matches in production-classified files were adjudicated by hand. Two were embedded test strings; the remaining 14 matches collapsed to 12 semantic construction paths in ten repositories. To make that filtering inspectable, the release includes the scanner, a privacy-safe 22-row frozen candidate projection, and an 11-row adjudication ledger with repository-relative files, line numbers, decisions, and pinned commits. We do not redistribute the clones, so the filtering decision is auditable even though acquisition is not fully reproducible.

\subsection{Low-load probes and released observations}
Fifteen sampled repositories advertised a remote endpoint. Our probes initialized the server, listed tools, called a guaranteed nonexistent name, supplied one wrong required-field type to a lexically read-only tool, and, for a conservative subset, sent a type-correct but out-of-domain sentinel. We neither exhausted quotas nor attempted to create state or trigger an outage. Ten endpoints initialized and yielded 21 verified failures. Their negotiated versions were 2025-03-26 (3), 2025-06-18 (3), and 2025-11-25 (4), as recorded in a released pseudonymous table. Five connection or initialization failures remain restricted and are not part of the tool-failure analysis. This low-risk design necessarily favors deterministic failures. It omits rate limits, genuine upstream faults, timeouts, partial commits, and uncertain idempotency---the very cases in which replay constraints matter most. Our $E/R$ result should be read within that boundary.

Python MCP SDK 1.26.0 serializes successful results through model dumping and records exceptions as normalized dictionaries. What we analyze is therefore the client's normalized view, not an untouched HTTP body or JSON-RPC frame. Server and tool identifiers are pseudonymous; endpoint URLs, headers, discovery schemas, timestamps, excluded success/search payloads, and unrestricted observations remain private.

We minimized before release rather than attempting to sanitize everything after the fact. Three service-specific observations were replaced by canonical minimal failures because the originals could retain irrelevant third-party fields. The builder, source hashes, redaction counts, and fail-closed privacy scanner are public; the restricted source records are not.

\subsection{Coding procedure}
The frozen codebook was applied independently to each of the six capabilities in both views. I performed this coding as the sole author. A previous AI-assisted pass remains in the artifact as a sensitivity record, but it is not a second human judgment and is not reported as inter-rater agreement. An archival reliability claim would require two trained independent coders and adjudication.

The borderline cases show why separate coding matters. S11 and S18 only say that lookup failed for the supplied test value, which establishes $D$ but no unique policy, cause, or target. S21 calls an identifier invalid, enough for $P$ and $C$, but never names the parameter and therefore misses $T$. These decisions can be checked directly in the row-level ledger.

\section{Exploratory Policy-Selection Stress Test}
\subsection{Conditions and evidence limits}
Each failure is replayed under four paired conditions:
\begin{description}[leftmargin=0pt,style=nextline,itemsep=2pt]
\item[Normalized.] The client-normalized observation alone.
\item[Code-only.] The observation plus a namespaced cause code, with no action field or recommended-action wording.
\item[Structured action.] The same code plus the preferred policy in JSON.
\item[Prose action.] The same cause and preferred policy in a sentence.
\end{description}
Only the code-only/normalized pair tests recovery inference without revealing the preferred answer. The other two conditions ask a different question. Structured action places the exact output label in JSON; prose action gives an imperative and a cause explanation. They expose different words as well as different formats, so their contrast is descriptive rather than a controlled format effect. High performance in either cell is evidence of compliance with an answer-bearing message, not independent recovery reasoning.

The test uses local Llama 3.2 3.2B (Q4\_K\_M) and Qwen 3.5 0.8B (Q8\_0) through Ollama 0.24.0. The runner checks the full immutable model digests recorded in the manifest before inference. Five stochastic replicates for every model, condition, and scenario produce $2\times4\times21\times5=\TrajectoryCount$ decisions; the extra replicates are repeated judgments, not new failures. Temperature is 0.7, with seeds paired across conditions. The model returns an \texttt{operation\_status} and a \texttt{recovery\_policy}; calling a verified failure \texttt{succeeded} counts as false success. No argument is actually repaired, credential acquired, alternative discovered, side effect checked, or task completed. The experiment is therefore about policy labels, not agent recovery.

When a model requests immediate replay, the simulator returns the same stored failure, for at most three attempts. The main descriptive outcome is exact agreement with the preferred policy. Coverage requires a valid policy other than \texttt{stop\_or\_escalate}; invalid responses count as neither coverage nor abstention. Because always stopping belongs to every non-harm set, that measure is only a safety check. We also record false success, invalid output, and needless replay. Results from two small models under one prompt should not be generalized to frontier or deployed agents.

\subsection{Descriptive analysis and deterministic baseline}
The paper reports exact counts for each model and condition, along with predeclared paired differences. Since the failures are not a probability sample and only ten servers contribute observations, an interval would not justify a population claim. Wilson and ten-server cluster-bootstrap calculations remain available as sensitivity analyses in the artifact. We avoid pooling across models and conditions because such a total would mix different systems and different amounts of disclosed information.

For comparison, we built a deterministic client. It first checks the ordinary completed-result shape---\texttt{resultType}, \texttt{content}, and the content blocks---then requires \texttt{isError:true} and validates the experimental schema, code, code--target pairing, policy, timing, replay rule, and side effects. A local registry maps the four controlled causes to host policy and adds a transient class for safe immediate or delayed replay. Anything malformed or unsupported stops safely. The prototype is an executable design baseline, not a production extension or a model leaderboard.

\section{Results}
\subsection{Specification and static code}
The specification check confirms the basic asymmetry. In the pinned 2026-07-28 schema, an ordinary completed \texttt{CallToolResult} has \texttt{resultType}, \texttt{isError}, content, and optional structured content, but no standard tool-failure code, repair target, policy, retry delay, or replay-safety field \cite{mcp_schema_2026}. Specialized \texttt{InputRequiredResult}/MRTR and typed protocol errors prevent any blanket claim that MCP has no structured recovery; they simply do not form a general taxonomy for completed failures. The SDK guidance places recovery hints in model-readable text, while issue 3003 treats schema governance on the error path as unresolved \cite{mcp_ts_errors,mcp_structured_error_issue}.

The repository audit tells a similar, though deliberately narrow, story. Each of the 12 adjudicated production error-construction paths combines \texttt{isError} with text. Seven produce prose alone; five serialize JSON or mixed JSON inside a text block. None provides a standard code or action. A separately audited repository does define a private structured recovery contract. That example demonstrates feasibility, but also fragmentation: a client cannot assume the contract elsewhere. The released ledger pins the repository and commit and records the safe relative files, line ranges, and fields.

\subsection{Joint capability profiles}
The 21 failures comprise ten nonexistent tools, seven invalid argument types, three invalid values, and one authentication failure. Eight normalize as protocol errors, ten as \texttt{isError} results, and three as success-shaped errors.

\begin{table}[t]
\centering
\caption{Separately coded capabilities for 21 observations. Coordinates are not cumulative.}
\label{tab:profile}
\small
\begin{tabularx}{\columnwidth}{@{}Xrr@{}}
\toprule Capability & Control & Content-assisted \\
\midrule
$D$ failure detection & \ControlDetect & \ContentDetect \\
$P$ coarse policy & \ControlPolicy & \ContentPolicy \\
$C$ cause/subtype & \ControlCause & \ContentCause \\
$T$ repair target & \ControlTarget & \ContentTarget \\
$E$ executable repair & \ControlExecutable & \ContentExecutable \\
$R$ replay constraints & \ControlReplay & \ContentReplay \\
\bottomrule
\end{tabularx}
\end{table}

Table~\ref{tab:profile} makes the information boundary visible. Typed fields identify \ControlDetect{} of the 21 records as failures. Eight \texttt{-32602} observations support a broad ``repair the request'' policy, but they cover five unknown tools and three invalid arguments; the code alone therefore identifies no specific cause. Once content is interpreted, all 21 failures are visible, \ContentPolicy{} support a coarse policy, \ContentCause{} identify a cause, and \ContentTarget{} name a target. Still, not one observation provides a complete replacement request, explicit retry timing, unchanged-request safety, or side-effect state.

With bits ordered $D/P/C/T/E/R$, the control profiles are $110000$ (\ControlJointDP), $100000$ (\ControlJointD), and $000000$ (\ControlJointNone). The content-assisted profiles are $111100$ (\ContentJointDPCT), $111000$ (\ContentJointDPC), and $100000$ (\ContentJointD). Both sets happen to form chains, with \ControlIncomparablePairs{} and \ContentIncomparablePairs{} incomparable pairs, so this sample does not exercise the partial order's main advantage. Nor should $E=R=0$ be overread. In the control view it follows from the missing standard fields; in the content view it describes only this safely inducible failure mix, not transient, rate-limit, upstream, timeout, partial-effect, or idempotency cases.

\subsection{Exploratory model-by-condition results}
\begin{table*}[t]
\centering
\caption{Exploratory policy-label selection; exact counts per 105 decisions. Strict is preferred-policy match; Cover is valid non-stop policy; Retry is wrong identical replay; Any FS is any-step false success; Extra is additional identical calls in the local simulator.}
\label{tab:agents}
\scriptsize
\begin{tabular}{@{}llrrrrrr@{}}
\toprule Model & Condition & Strict & Cover & Retry & Invalid & Any FS & Extra \\
\midrule
Llama 3.2 & Normalized & \LlamaNormalizedStrictCount & \LlamaNormalizedCoverageCount & \LlamaNormalizedRetryCount & \LlamaNormalizedInvalidCount & \LlamaNormalizedAnyFalseSuccessCount & \LlamaNormalizedAdditionalCalls \\
Llama 3.2 & Code-only & \LlamaCodeStrictCount & \LlamaCodeCoverageCount & \LlamaCodeRetryCount & \LlamaCodeInvalidCount & \LlamaCodeAnyFalseSuccessCount & \LlamaCodeAdditionalCalls \\
Llama 3.2 & Structured & \LlamaStructuredStrictCount & \LlamaStructuredCoverageCount & \LlamaStructuredRetryCount & \LlamaStructuredInvalidCount & \LlamaStructuredAnyFalseSuccessCount & \LlamaStructuredAdditionalCalls \\
Llama 3.2 & Prose & \LlamaProseStrictCount & \LlamaProseCoverageCount & \LlamaProseRetryCount & \LlamaProseInvalidCount & \LlamaProseAnyFalseSuccessCount & \LlamaProseAdditionalCalls \\
Qwen 0.8B & Normalized & \QwenNormalizedStrictCount & \QwenNormalizedCoverageCount & \QwenNormalizedRetryCount & \QwenNormalizedInvalidCount & \QwenNormalizedAnyFalseSuccessCount & \QwenNormalizedAdditionalCalls \\
Qwen 0.8B & Code-only & \QwenCodeStrictCount & \QwenCodeCoverageCount & \QwenCodeRetryCount & \QwenCodeInvalidCount & \QwenCodeAnyFalseSuccessCount & \QwenCodeAdditionalCalls \\
Qwen 0.8B & Structured & \QwenStructuredStrictCount & \QwenStructuredCoverageCount & \QwenStructuredRetryCount & \QwenStructuredInvalidCount & \QwenStructuredAnyFalseSuccessCount & \QwenStructuredAdditionalCalls \\
Qwen 0.8B & Prose & \QwenProseStrictCount & \QwenProseCoverageCount & \QwenProseRetryCount & \QwenProseInvalidCount & \QwenProseAnyFalseSuccessCount & \QwenProseAdditionalCalls \\
\bottomrule
\end{tabular}
\end{table*}

In the only answer-free comparison, adding a symbolic cause code changes strict matches from 42 to 39 for Llama and from 6 to 5 for Qwen. The taxonomy label does not help either model in this setup. When the message discloses the action, structured JSON yields 95 and 39 strict matches, while prose yields 105 and 30. Those cells mainly measure whether the model follows an answer-bearing instruction; their lexical differences rule out a clean format comparison. An always-stop baseline is non-harmful in all 21 scenarios but achieves no strict match or coverage. The code-registry and explicit-policy baselines reach 21/21 by construction.

Needless replay is concentrated in the smaller model. Llama repeats the failed request once in each answer-free condition and never does so when the action is supplied. Qwen repeats it 24, 24, 26, and 23 times across normalized, code-only, structured, and prose conditions; because the loop permits further attempts, those choices create 48, 48, 55, and 42 additional calls. Llama adds one, one, zero, and zero. Qwen also accounts for every terminal false-success judgment: 4, 1, 1, and 0 by condition. These figures diagnose this controller and these models. They do not measure load amplification on a degraded service.

\subsection{Executable conformance result}
The deterministic prototype handles all 21 counterfactual controls as designed. Registry and explicit-action modes select both the preferred and a non-harmful policy in every case. Two synthetic transient cases cover safe immediate replay and delayed retry. Seventeen malformed or unsafe inputs must fail closed with an exact reason and stop/escalate; these include missing base content, invalid content shape, three code--target mismatches, and unsafe reachable replay. The suite establishes that the design is executable inside one implementation. It does not establish interoperability or protocol security.

\section{An Evidence-Motivated Two-Plane Candidate}
The evidence points to a modest design rule: do not ask one error string to serve both software and people. Building on earlier structured-recovery proposals, we place validated controls in a \emph{control plane} for the host and leave the explanation in content for models and people:
\Needspace{0.40\textheight}
\begin{lstlisting}[language={},caption={Illustrative experimental control envelope.},label={lst:envelope}]
{
  "resultType": "complete",
  "isError": true,
  "error": {
    "schema": "urn:example:mcp-error:1",
    "code": "example.mcp.arguments.invalid_type",
    "target": {"kind": "parameter",
               "pointer": "/walletAddress"},
    "policy": "revise_arguments",
    "retry": {"sameRequest": "unsafe"},
    "sideEffects": "none_committed"
  },
  "content": [{"type": "text",
    "text": "walletAddress must be a string."}]
}
\end{lstlisting}

The host first validates the ordinary result, then checks the experimental schema, code--target pairing, policy, replay constraints, and side effects. Only after those checks does it apply local authorization, confirmations, retry budgets, backoff, or circuit breakers. The server is still making assertions: its code, policy, replay claim, and side-effect claim are not automatically trustworthy. A host must compare them with request identity, local state, and authority. Unknown codes and unavailable capabilities stop safely, while the explanation remains available. The \texttt{example} namespace is intentionally non-official.

Where such a control belongs in MCP is still an open design question. A reserved member would need capability negotiation and independent versioning so that a successful tool's \texttt{outputSchema} cannot quietly redefine failure semantics. Older clients could continue reading \texttt{isError} and content; aware hosts could inspect validated controls before involving a model. Our 21-case implementation goes no further than internal executability. Independent implementations, downgrade behavior, namespace collision, malicious assertions, extension negotiation, SDK validation, and standardization remain untested.

\section{Threats, Ethics, and Reproducibility}
\paragraph{Sampling and representation.} The registry is still a preview, and our sample is not a census. Reachability and low-risk inducibility both select which failures enter the study. Transient, rate-limit, upstream, timeout, partial-commit, and idempotency cases are absent, so the $E/R$ result cannot speak for them. The lexical audit exhausts one high-precision candidate class rather than every possible error path, and SDK normalization changes fields and exception shapes. These data illustrate a protocol boundary; they do not estimate prevalence or preserve raw-wire behavior.

\paragraph{Coding and oracle.} The profile is rule-based and coded by one author. Public rationales make each decision inspectable, but they are no substitute for independent human coding. Because every non-harm set permits stopping, the 100\% always-stop baseline cannot be interpreted as progress. Preferred policies also remain judgments: a real task may allow clarification, discovery, or user-controlled authorization outside our vocabulary.

\paragraph{Models and inference.} Two small local models, one prompt, and 21 scenarios cannot characterize deployed agents, especially when two treatments reveal the answer. Qwen accounts for nearly all wrong retries and invalid responses, which may simply reflect model scale. Prose and structure also differ in wording and explanation, not only serialization. We had neither frontier-model credentials nor budget and did not invent a proxy for them. Exact cells lead the analysis; sensitivity intervals remain in the artifact because a non-probability corpus and ten server clusters offer little inferential reach.

\paragraph{Behavioral scope.} The simulator records policy labels, not repaired requests, downstream completion, or side effects. Replaying produces the same stored response. There is no backoff, shared load, service recovery, or real transient outage, so the additional calls are properties of this local controller rather than a degradation result.

\paragraph{Ethics and privacy.} Probes were low volume and limited to initialization, nonexistent names, malformed arguments on conservatively selected read-only tools, and sentinel values. We did not obtain maintainer consent, conduct a jurisdiction-specific terms review, seek an IRB determination, or coordinate disclosure. No production calls were made during the later revisions. Endpoint and discovery payloads are absent from the release, and unrestricted originals stay in private work storage. An earlier unsafe draft bundle was quarantined and is not distributed.

A larger IMC/PAM-style collection should begin with an identifying User-Agent linked to a study page, an opt-out route, community notice, an IRB determination or documented exemption, and coordinated disclosure of success-shaped correctness bugs. None of those safeguards should be inferred from this pilot; they are requirements for follow-on work.

\paragraph{Artifact.} The privacy-safe package reproduces the \emph{derived analysis}. It contains frozen sampling metadata, exact draw and replacement reconstruction, pseudonymous protocol versions, minimized observations, record-level rationales, non-harm sets, the lexical scanner and frozen candidates, static provenance, \TrajectoryCount{} parsed decisions, analysis and conformance code, a runtime manifest, and hashes. The documented commands use only the Python standard library and run relative to the extracted package. Raw model responses were not retained, so a third party cannot re-audit parser and invalid-output decisions; a future run should preserve those responses and full row-level digests. Clones, the raw registry payload, acquisition traffic, and unrestricted endpoint data remain outside the package.

\section{Discussion and Conclusion}
Calling an error ``machine-actionable'' hides several different questions. Can software see the failure? Can it choose a kind of response? Does it know what broke, what to change, and whether replay is safe? The six-part profile keeps those questions separate. Within our completed-result-only boundary, deterministic software often sees the failure and sometimes identifies a broad response. It usually lacks the cause, target, executable repair, or replay conditions unless it reads server-specific prose or consults other state. Prose is useful evidence, but relying on it puts a semantic interpreter---and its failure modes---inside the recovery path.

The model test supports only a modest warning. Showing these two small models a taxonomy name did not improve policy selection; showing them the answer mostly tested whether they would follow it. Their needless retries and false-success judgments are reasons to keep failure belief separate from recovery policy, not estimates of how deployed agents behave or how much load retries create during an outage.

The two-plane example is best understood as a conformance target grounded in the audit, not as the first proposal for structured MCP recovery. At minimum, such a control plane needs a stable cause code, policy compatibility, an optional target, and explicit assertions about replay and side effects. Explanations can remain readable by people and models. Our client branches deterministically and fails closed on controlled cases. Whether independent implementations interoperate---and whether the assertions survive adversarial conditions---is still unknown.

The paper's answer is not simply yes or no. For ordinary completed tool errors, MCP standardizes observability more strongly than recovery semantics, with important exceptions in typed specialized workflows. Software can often branch on \emph{failure} and sometimes on a broad policy class. In this sample, however, concrete autonomous recovery remains dependent on prose or information outside the result. Testing how far that conclusion travels will require independent human coding, raw-wire captures across SDKs, the missing failure classes, end-to-end recovery experiments, and independent interoperability and adversarial trials.

\section*{Acknowledgment}
Generative AI tools (OpenAI Codex and Anthropic Claude) assisted with literature discovery, code review, adversarial manuscript review, drafting, and prose editing. The author selected the research question and methods, verified the cited sources and derived results, made the scientific judgments, and accepts responsibility for the paper.

\bibliographystyle{IEEEtran}
\bibliography{references}
\end{document}